\documentclass[prl,twocolumn,aps,showpacs,superscriptaddress]{revtex4-1}

\usepackage{graphicx}

\begin{document}

\title{Topological superconductivity at the edge of transition metal dichalcogenides}

\author{Gang Xu$^{1,2}$, Jing Wang$^1$, Binghai Yan$^{3,4}$ and Xiao-Liang Qi}

\affiliation{Department of Physics, McCullough Building, Stanford University, Stanford, CA 94305-4045, USA\\
$^2$Beijing National Laboratory for Condensed Matter Physics,
and Institute of Physics, Chinese Academy of Sciences, Beijing 100190, China\\
$^3$ Max Planck Institute for Chemical Physics of Solids, D-01187 Dresden, Germany \\
$^4$ Max Planck Institute for Physics of Complex Systems, D-01187 Dresden, Germany}

\date{\today}

\begin{abstract}
Time-reversal breaking topological superconductors are new states of matter which can support Majorana zero modes at the edge. In this paper, we propose a new realization of one-dimensional topological superconductivity and Majorana zero modes. The proposed system consists of a monolayer of transition metal dichalcogenides MX$_2$ (M=Mo, W; X=S, Se) on top of a superconducting substrate. Based on first-principles calculations, we show that a zigzag edge of the monolayer MX$_2$ terminated by metal atom M has edge states with strong spin-orbit coupling and spontaneous magnetization. By proximity coupling with a superconducting substrate, topological superconductivity can be induced at such an edge. We propose NbS$_2$ as a natural choice of substrate, and estimate the proximity induced superconducting gap based on first-principles calculation and low energy effective model. As an experimental consequence of our theory, we predict that Majorana zero modes can be detected at the 120$^\circ$ corner of a MX$_2$ flake in proximity with a superconducting substrate.
\end{abstract}

 \maketitle

{\it Introduction.--}In recent years, topological insulators (TI) and topological superconductors (TSC) have been discovered in different dimensions and symmetry classes\cite{hasan2010,moore2010,qi2011rmp}. In particular, time-reversal breaking TSC are proposed in one or two dimensions (1D or 2D)\cite{kitaev2001,read2000}. One important motivation to study TSC is that they can realize Majorana zero modes at the edge (for 1D) or vortex core (for 2D). A Majorana zero mode is half of an ordinary fermion zero mode, which carries nonlocal degrees of freedom and makes the TSC system a candidate for topological quantum computation\cite{nayak2008}. Recently, many theoretical proposals have been made for the realization of
Majorana zero modes~\cite{fu2008,sau2010,lutchyn2010c,oreg2010,alicea2010,potter2011,lee2009,Duckheim2011,Weng2011,Chung2011}. (For a recent review, see Ref.\cite{alicea2012}.) In particular, it has been proposed that a nanowire in proximity to an $s$-wave superconductor, with spin-orbit coupling (SOC) and magnetic field, can realize the 1D TSC phase~\cite{lutchyn2010c,oreg2010}.
Recently, significant experimental progress has been made towards the realization of this proposal~\cite{mourik2012,Deng2012,Das2012,Rokhinson2012,finck2013,rodrigo2012}, although the interpretation of the experimental results has not been completely settled.

In this paper, we propose to realize topological superconductivity and Majorana zero modes in a new family of materials, the monolayer transition metal dichalcogenides MX$_2$ (M=Mo, W; X=S, Se) in proximity with a conventional superconductor. Based on {\it ab initio} calculations, we predict that a particular edge of this material (as will be explained below) has 1D edge states with strong Rashba SOC and spontaneous magnetization. At a suitable Fermi level, the interplay of SOC and magnetization creates a 1D system with a single pair of Fermi points, so that the proximity effect with an $s$-wave superconductor will induce a TSC. Compared to previous nanowire proposals, our proposal has several advantages. Firstly, this proposal is simpler to realize since it does not require creating a nanowire. The 1D state is given automatically at the edge of single layer MX$_2$. Secondly, due to the spontaneous magnetization of the edge state, no external magnetic field is needed. Thirdly, superconductors NbS$_2$ and NbSe$_2$ (with p6$_3$/mmc symmetry) have the same structure as the MX$_2$ semiconductors and similar lattice constants, which indicates that a relatively strong proximity coupling could be induced between them.  We investigate different materials in this family and find that the optimal combination is MoS$_2$ monolayer on the substrate of NbS$_2$ superconductor. To verify our proposal experimentally, we propose that a single Majorana zero mode can be found at a 120$^\circ$ corner of a MoS$_2$ ribbon.

\begin{figure}[tbp]
\includegraphics[width=3.0in,scale=0.55,angle=270]{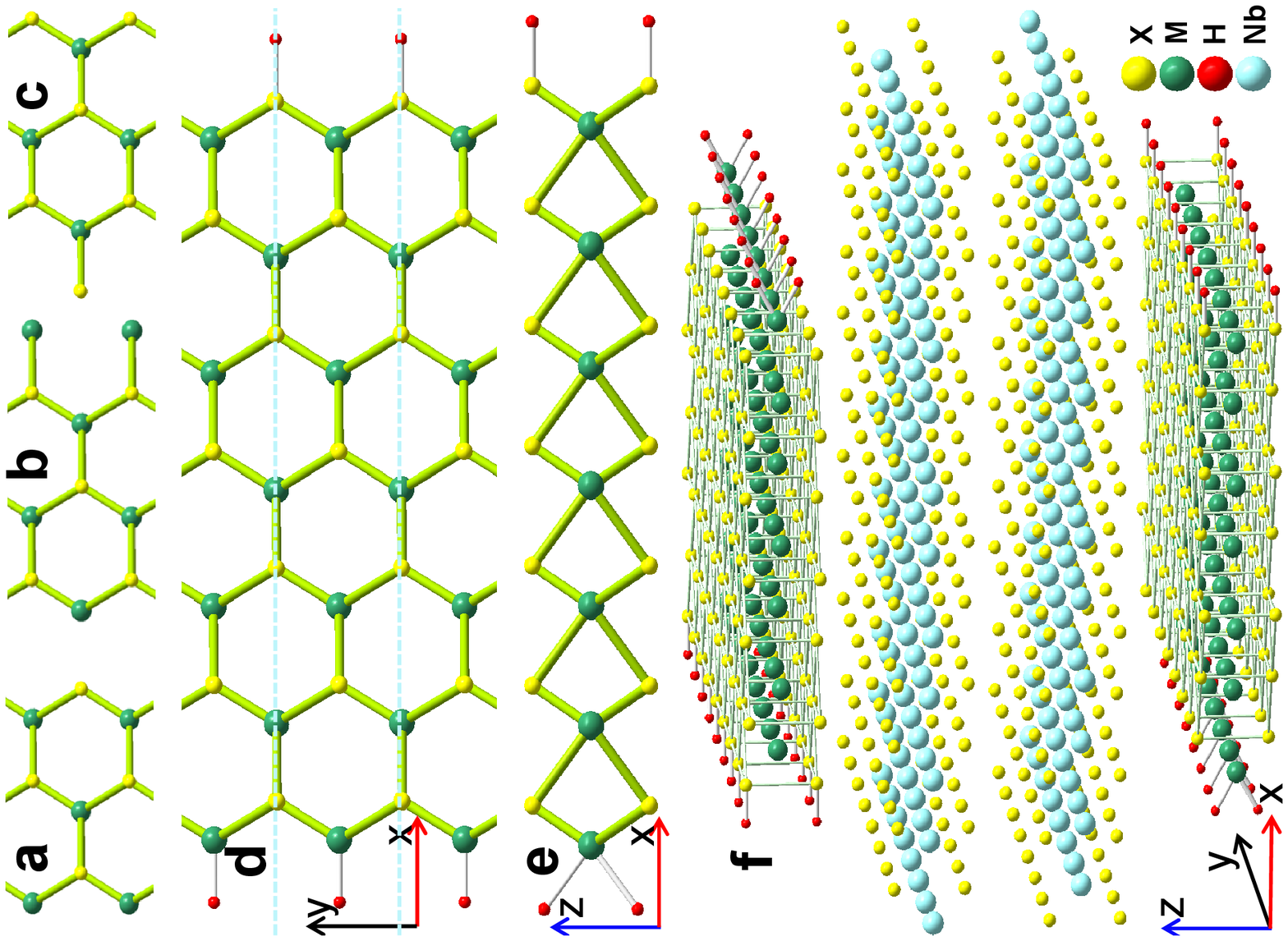}
\caption{(Color online)   (a), (b) and (c): zigzag MX$_2$ nanoribbons with three different boundary conditions zz-MX, zz-M and zz-X,  respectively.
  (d) and (e) Top and side view of a monolayer zz-MX ribbon with saturated hydrogen configuration.
 (f) The heterostructure of a MX$_2$ nanoribbon on NbS$_2$ substrate.}
\end{figure}

{\it First-principles calculation on MX$_2$ nanoribbons.--}The existence of 1D metallic edge states on the S-terminated zigzag edge MoS$_2$ has been well confirmed by experiments and density functional theory (DFT) calculations~\cite{Bollinger2001,YFLi2008,Vojvodic2009,Pan2012,Kou2012}. Encouraged by the successful experimental observation, an explosion of theoretical studies on MoS$_2$ has occurred, leading to discoveries of interesting electronic and magnetic properties on several edge structures.
Recently, Pan \emph{et al} find that the ferromagnetic states of the MoS$_2$ zigzag nanoribbons can be enhanced by hydrogen saturation~\cite{Pan2012}, as well as in-plane electric field~\cite{Kou2012}. On the other hand, due to the absence of inversion symmetry, Rashba type SOC naturally exists in the edge states. The interplay of magnetization and Rashba SOC lifts the spin degeneracy and can lead to a single pair of Fermi points. Therefore, the zigzag nanoribbons of monolayer MX$_2$ may become ideal candidates of 1D TSC when it is in proximity with an $s$-wave superconductor. In the following we will investigate several different edge conditions by {\it ab initio} calculation to search for the ideal edge states for our purpose.

Our first-principles calculations are carried out by the Vienna Ab-initio Simulation Package (VASP)~\cite{Kresse1993,Kresse1999} within the framework
density functional theory~\cite{Hohenberg1964}. We use the generalized gradient approximation (GGA) of Perdew-Burke-Ernzerhof type~\cite{Perdew1996} for the exchange-correlation potential. The kinetic energy cutoff is fixed to 450 eV. 2$\times$16$\times$1 k-mesh is used for monolayer nanoribbon calculations. The lattice constants and atomic positions are fully optimized, in which the maximal force at an ion is smaller than 0.01 eV/\AA.
Based on previous studies~\cite{Pan2012}, we consider three types of nanoribbons as is shown in Fig. 1(a-c). The first one (Fig. 1 (a)) labeled by zz-MX has two edges terminated by M and X atoms respectively. The second one (Fig. 1 (b)) labeled by zz-M is terminated by M atoms for both edges. The third one (Fig 1. (c)) labeled by zz-X is terminated by X atoms for both edges. In order to simulate the more realistic and stable boundaries~\cite{Lauritsen2003}, hydrogen saturation are considered by adding two hydrogen atoms to the M atoms and one hydrogen atom to the X atoms at the edges, as is shown in Fig. 1(d) and Fig. 1(e).

\begin{figure}[tbp]
\includegraphics[clip,scale=0.35, angle=270]{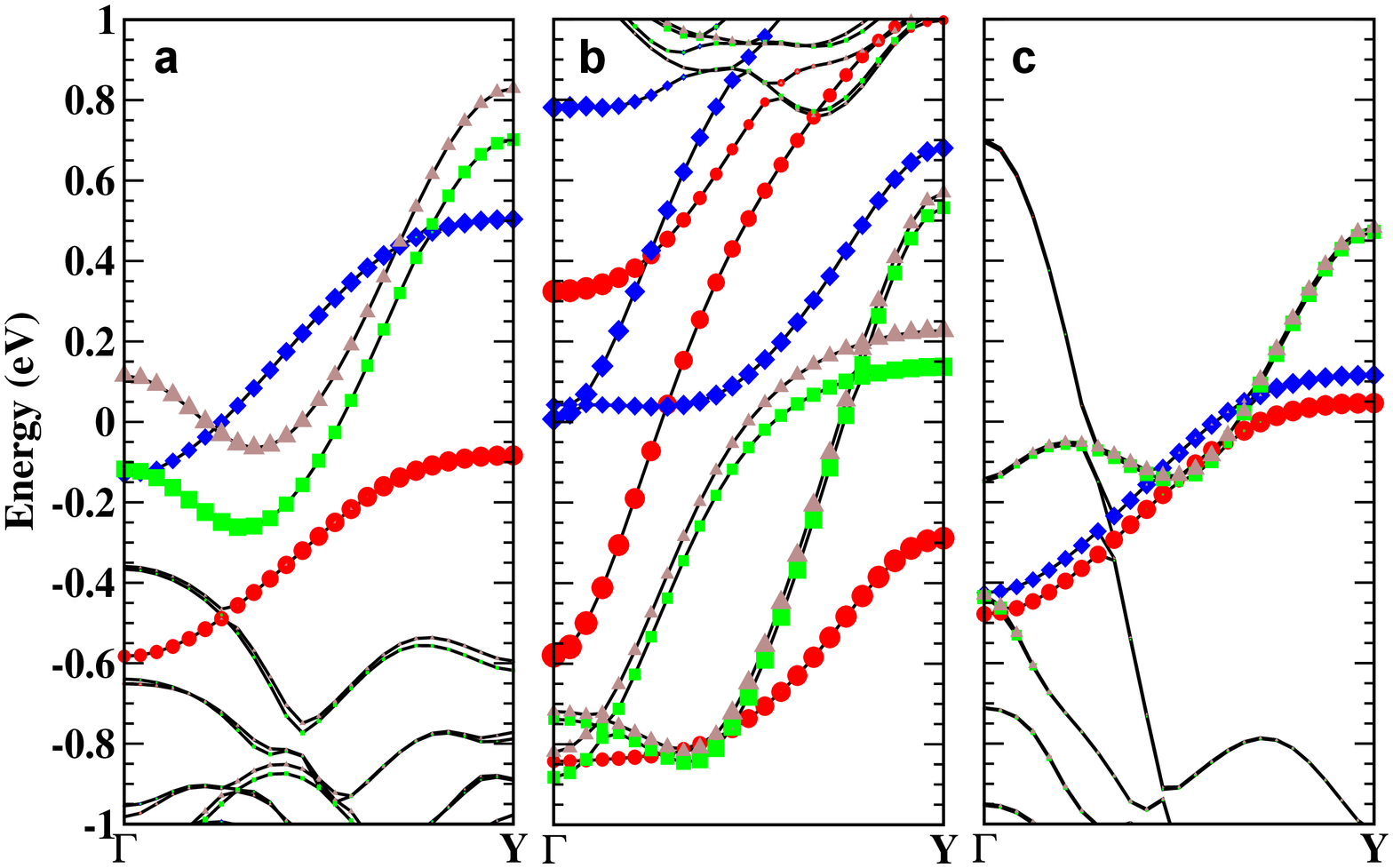}
\caption{(Color online) {Spin polarized band structures
  without SOC for zz-MoS (a), zz-Mo (b) and zz-S (c).}
  Red circles and Blue diamonds denote the projections
  to the up and down spin $d$-orbitals of the Mo atoms on right edges shown in Fig. 1.
  The projections to Mo on left edges are indicated
  as green squares and brown triangles, respectively.
  The Fermi level is defined at 0 eV}
\end{figure}

We start by studying the band structure of free-standing nanoribbons. First, we carry out spin-polarized calculations without including SOC, to check the electronic
and magnetic properties of all three types of zigzag nanoribbons. In this case, all the transition metal dichalcogenides are very similar. Here we show band structures of MoS$_2$ in Fig. 2 as a representative. For all three types of nanoribbons one can see that edge states are metallic and spin-polarized,
which are clearly distinct from the non-magnetic insulating bulk states of the monolayer MoS$_2$. There is only one band per spin direction crossing the Fermi level for each edge of zz-MoS, which is dominated by Mo-$d$ orbitals. The band from left and right edge has a spin polarization of 0.61 $\mu_B$ and 0.27 $\mu_B$ per Mo atom, respectively. Similarly, the metallic bands of zz-Mo ribbon are also mainly contributed by Mo-$d$ orbitals. However, due to many bands presenting around the Fermi level ($E_f$), it's too complicated to get a single pair of Fermi points with opposite spin moments. For zz-S case, there is not only Mo $d$ orbital bands but also
a spin degenerate $p$ orbital band of S crossing the $E_f$. Due to the insulating block between two edges, the two edge states can be considered as decoupled.
Therefore the band structures in Fig. 2 suggests that both edges of zz-MoS ribbon or the right edge of zz-S ribbon are good candidates of simple 1D channels. However, the magnetic splitting for zz-S ribbon is always smaller than 0.1 $\mu_B/M$ for all four transition metal dichalcogenides MX$_2$, and will be further reduced when SOC effect are considered. Therefore, we will focus on the zz-MX ribbon in the following, which exhibits sufficient magnetization and a simple edge band structure.

Now we carry fully relativistic calculations, to study the influence of SOC
on the electronic and magnetic properties. We calculate three different configurations with moments along x, y (in-plane) and z directions (out-of-plane) for all four transition metal dichalcogenides, and find that the two in-plane cases have the same energy gains, magnetic moments and the electronic structures, while there is strong anisotropy between the in-plane and out-plane directions. For MoS$_2$, the in-plane polarization is $0.4$ meV lower in energy than the out-of-plane configuration. In contrast, for WS$_2$ the in-plane energy is $7.6$ meV higher than that of out-of-plane. For the diselenides MoSe$_2$ and WSe$_2$, the system always prefer an out-of-plane magnetization, and the calculation never converges to an in-plane configurations. Such difference between sulfides and selenides can be understood as a consequence of the competition between ferromagnetism and SOC.

\begin{figure}[tbp]
\includegraphics[clip,scale=0.33,angle=270]{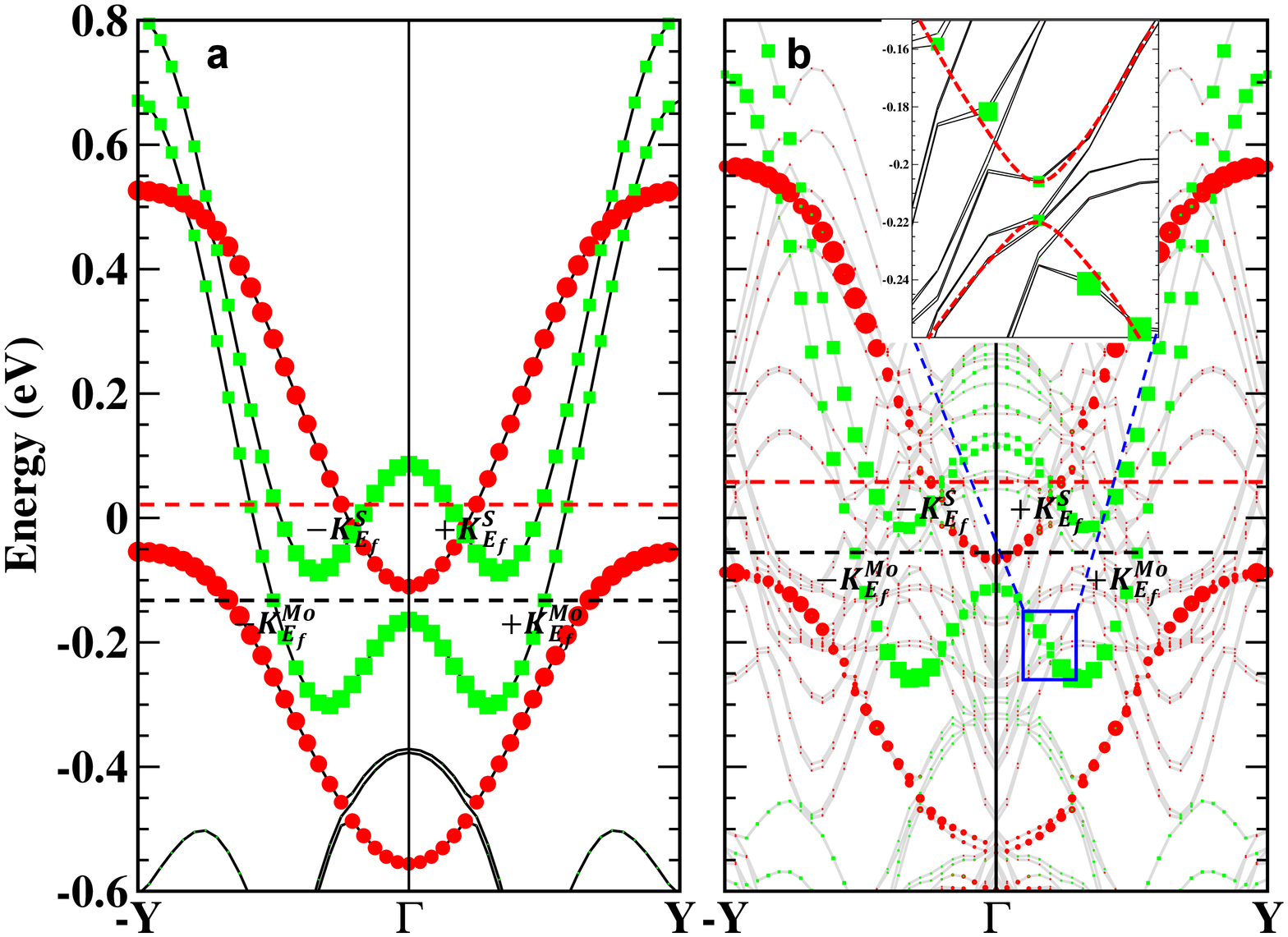}
\caption{(Color online) {Electronic structures with Rashba SOC effect
  of zz-MoS nanoribbon (a) and the zz-MoS on NbS$_2$ substrate (b).}
  The size of red circles and green squares mean the projection to Mo located on right and left edges respectively.
  The red and black dash mean the suitable Fermi level ($E_f$) for S-terminated and Mo-terminated edge respectively.
 $\pm$K$^{S}_{E_f}$ and $\pm$K$^{Mo}_{E_f}$ are defined on each $E_f$ for comparing their spin moments.
  The inset of (b) is the zoom-in of the blue box,
  in which the the red dashes are the schematic curves of the hybridization splitting between MoS$_2$ and NbS$_2$ bands.
  }
\end{figure}

As will be analyzed more carefully in the later part of the article, the free-standing monolayer of MX$_2$ has a reflection symmetry according to the xy plane, which preserves the M layer and exchanges the two X layers. To preserve this symmetry, the Rashba spin splitting has to occur along $z$ direction, where $z$ spin components of two states  at momenta $\mathbf{k}$ and $\mathbf{-k}$ along the edge should be opposite. In contrast to the out-of-plane magnetization that does not open a gap, the in-plane magnetization breaks the reflection symmetry and thus leads to a gap opening at $\Gamma$ point.
For the search of topological superconductivity, it is necessary to gap the bands near $\Gamma$ point, in order for a single pair of Fermi points to occur. Therefore MoS$_2$ is the best candidate material, which has in-plane magnetization as the lowest energy configuration. In principle it is possible to have WS$_2$ and apply an in-plane magnetic field to rotate the magnetization to in-plane directions, but this only applies if the superconductor that is in proximity with the nanoribbon has a large enough pairing gap. The band structure of MoS$_2$ with SOC is shown in Fig. 3(a), from which one can see that
both Mo-terminated and S-terminated edges can have a single pair of Fermi points in a window of Fermi level position. For Mo-terminated edge, the window is about $75$ meV, and for S-terminated edge it is much larger. (To pin the Fermi level in this window, a top gate generically need to be applied. The gate control of Fermi level should not be too difficult since the edge state is one-dimensional. )
We calculate the spin components for typical Fermi level positions $\pm$K$^{Mo}_{E_f}$ and $\pm$K$^{S}_{E_f}$ (marked in Fig. 3(a))
and list them in Table I. For the S-terminated edge, less than 1\% spin components remain antiparallel as a result of strong magnetic splitting and weak SOC of the S-$p$ orbitals, which makes it hard to induce superconductivity by proximity effect at this edge. In contrast, for the Mo-terminated edge, besides the parallel moments along x-direction yielded by ferromagnetic requirement, about 35\% spin components along z-direction are arranged antiparallel as a results of Rashba effect.
These sizable opposite spin components would make the proximity effect with an $s$-wave superconductor much easier than the S-terminated edge. In the following, we will study the proximity effect of this Mo-terminated edge with superconductors.

\begin{table}
\caption{Calculated spin moments at $\pm$K$^{Mo}_{E_f}$ and $\pm$K$^{S}_{E_f}$ for MoS$_2$ and MoS$_2$ on NbS$_2$ substrate.}
\scriptsize{
\begin{tabular}{l|cc|cc|cc|cc}
\hline
\hline
\multicolumn{1}{c|}{\quad}         &\multicolumn{4}{c|}{zz-MoS$_2$} &\multicolumn{4}{c}{zz-MoS$_2$ on NbS$_2$}\\
\hline
\multicolumn{1}{c|}{\quad}         &\multicolumn{2}{c|}{Mo edge} &\multicolumn{2}{c|}{S edge}
                                   &\multicolumn{2}{c|}{Mo edge} &\multicolumn{2}{c}{S edge}\\
\hline
\multicolumn{1}{c|}{M($\mu_B$/Mo)} &\multicolumn{2}{c|}{0.29}       &\multicolumn{2}{c|}{0.61}
                                   &\multicolumn{2}{c|}{0.32}       &\multicolumn{2}{c}{0.65}\\
\hline
\quad   &-K$^{Mo}_{E_f}$ &+K$^{Mo}_{E_f}$ &-K$^{S}_{E_f}$ &+K$^{S}_{E_f}$ &-K$^{Mo}_{E_f}$ &+K$^{Mo}_{E_f}$ &-K$^{S}_{E_f}$ &+K$^{S}_{E_f}$ \\
$\langle S\rangle$      &0.828    &0.828  &0.742    &0.742    &0.829    &0.829  &0.745    &0.745\\
$\langle S_x\rangle$    &0.772    &0.772  &0.740    &0.740    &0.784    &0.783  &0.737    &0.737\\
$\langle S_y\rangle$    &0.000    &0.000  &0.000    &0.000    &0.000    &0.000  &0.000    &0.000\\
$\langle S_z\rangle$    &-0.286   &0.286  &-0.003   &0.003    &-0.269   &0.272  &-0.030   &0.030\\
\hline
\hline
\end{tabular}
}
\label{table:BNA}
\end{table}

To induce the TSC phase, the MoS$_2$ nanoribbon needs to be coupled with an $s$-wave superconductor. A natural choice is NbS$_2$ (with T$_c$ = 6.5 $K$), which has the same structure as MoS$_2$. The in-plane lattice of NbS$_2$ is 3.364 \AA~\cite{Jones1972}, a little lager than that of MoS$_2$ 3.188 \AA~\cite{Boker2001}. Due to the weak van der Waals interactions between neighboring layers, it should be possible to fabricate this device experimentally even with 5\% lattice mismatching. To eliminate the artificial electric field along z direction, we study a centrosymmetric heterostructure of MoS$_2$ nanoribbon and NbS$_2$, as is shown in Fig. 1 (f). Periodic boundary condition is taken for the NbS$_2$ layers in the $x,y$ directions. Comparing to the free-standing MoS$_2$ ribbon, the energy of the in-plane polarization is enhanced to $3.7$ meV lower than the out-of-plane configuration for such heterostructure. The calculated band structures are shown in Fig. 3(b). Compared to that of the free-standing MoS$_2$ ribbon in Fig. 3 (a), the edge state band structure changes very little. This result confirms that the coupling with metallic NbS$_2$ does not change the edge state bandstructure qualitatively, due to the weak Van der Waals interactions. The main difference of Fig. 3 (b) and Fig. 3 (a) is that a series of bands of NbS$_2$ appear from -0.5 $\sim$ 0.7eV, which cross the MoS$_2$'s edge states at Fermi level with very small hybridization shown in the inset of Fig. 3 (b). This hybridization, although small, is essential for inducing superconducting proximity effect.

{\it Effective model description.--}To understand the superconducting proximity effect more quantitatively, we introduce an effective model to describe the Mo-terminated zigzag edge coupled with the superconductor substrate. We start by a symmetry analysis of the nanoribbon. MoS$_2$ has a layered crystal structure with space group p6$_3$/mmc. The zigzag edge along $y$-direction breaks the symmetry to $C_{2v}$, which consists of a two-fold rotation $C_2$ around $x$-axis and a mirror operation $\sigma_d:z\rightarrow-z$ where $z$ is normal to the film. In the nonmagnetic state, the edge state Hamiltonian $H_0(k)$ is symmetric under $C_2,\sigma_d$ and time reversal $T$. In these symmetry operations the momentum and spin transform as follows:
\begin{eqnarray}\label{symmetry}
C_2: & k\rightarrow -k,\ \ \sigma_x\rightarrow\sigma_x,\ \ \sigma_y\rightarrow-\sigma_y,\ \ \sigma_z\rightarrow-\sigma_z,\nonumber
\\
\sigma_d: & k\rightarrow k,\ \ \sigma_x\rightarrow-\sigma_x,\ \ \sigma_y\rightarrow-\sigma_y,\ \ \sigma_z\rightarrow\sigma_z,\nonumber\\
T: &k\rightarrow -k,\ \ \sigma_x\rightarrow-\sigma_x,\ \ \sigma_y\rightarrow-\sigma_y,\ \ \sigma_z\rightarrow-\sigma_z,
\end{eqnarray}
These constraints together requires $H_0$ to have the form of $H_0(k)=\epsilon_0(k)+\epsilon_z(k)\sigma_z$. Adding the exchange splitting term due to the magnetic moment along $x$-axis, we obtain the following effective Hamiltonian up to fourth order of the momentum:
\begin{equation}
\mathcal{H}_{1D}(k) = \epsilon_0(k)+(\beta_1k+\beta_2k^3)\sigma_z+\gamma m\sigma_x,
\end{equation}
where $\epsilon_0(k)=\alpha_1k^2+\alpha_2k^4$. By fitting the energy band dispersion with the result of {\it ab initio} calculation, we obtain the parameters $\alpha_1 = -88.25$~eV$\cdot$\AA$^2$, $\alpha_2=1.61 \times 10^4$~eV$\cdot$\AA$^4$, $\beta_1=3.205$~eV$\cdot$\AA, $\beta_2=-7.232\times10^2$~eV$\cdot$\AA$^3$, $\gamma=0.43$~eV/$\mu_B$ and $m=0.29\mu_B$.

\begin{figure}[tbp]
\includegraphics[clip,scale=0.27,angle=270]{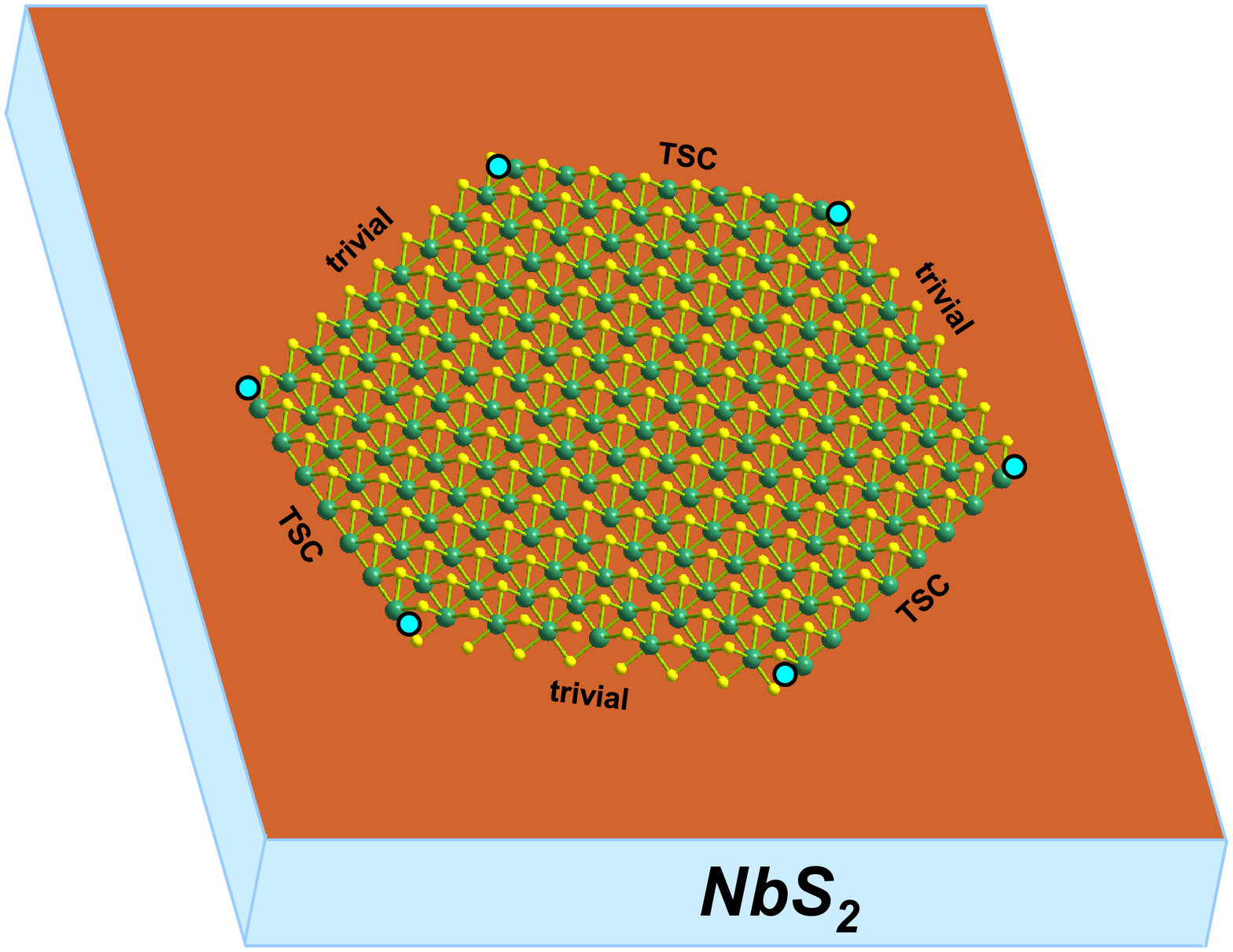}
\caption{(Color online) Schematic picture of the proposed device with a MoS$_2$ flake on the NbS$_2$ substrate. The Mo-terminated edges are TSC and the S-terminated edges are trivial, so that each 120$^\circ$ corner traps a Majorana zero mode (blue disk). }
\end{figure}

Now we consider the proximity effect between this 1D edge state and a 2D $s$-wave superconductor (SC). The effective Hamiltonian of the coupled system can be written as
\begin{equation}
\mathcal{H} = \mathcal{H}_{\text{SC}}+\mathcal{H}_{\text{1D}}+\mathcal{H}_{t},
\end{equation}
with $\mathcal{H}_{\text{SC}} = \sum_{\mathbf{k},\sigma}(\epsilon'_{\mathbf{k}}-\mu')c_{\mathbf{k}\sigma}^{\dag}c_{\mathbf{k}\sigma}
+\sum_{\mathbf{k}}(\Delta c^\dag_{\mathbf{k}\uparrow}c^\dag_{-\mathbf{k}\downarrow}+\text{H.c.})$ the Hamiltonian of an $s$-wave superconductor, $\mathcal{H}_{\text{1D}} = \sum_{k,\sigma,\sigma'} f^{\dag}_{k\sigma}(\epsilon_{k}^{\sigma\sigma'}-\mu\delta_{\sigma\sigma'})f_{k\sigma'}$ that of the edge state, and $\mathcal{H}_t = \sum_{\mathbf{k},\sigma}(t_{\mathbf{k}}f^{\dag}_{k\sigma}c_{\mathbf{k}\sigma}+\text{H.c.})$ the coupling term. Note that the spin quantization axis is along the $z$ direction and $k$ is the zigzag edge projection of $\mathbf{k}=(k_x,k)$. This model provides a minimal model for the SC proximity effect. For simplicity we only consider the spin-conserving hopping with constant amplitude $t_{\mathbf{k}}=t$. From the first-principles calculations shown in Fig. 3 (b), one can estimate the hopping amplitude from the avoid crossing between edge state and NbS$_2$ bulk states as $t\approx 22.1$~meV.

For our purpose we are interested in the case when the edge state Fermi level only crosses the lower band, as is shown in Fig. 3. The pairing at the Fermi surface is given by $\langle \xi_{-k}^-\xi_{k}^-\rangle$ with $\xi_{k}^-=a_{k}f_{k\uparrow}+b_{k}f_{k\downarrow}$ the annihilation operator in the lower band. $a_{k}/b_{k}=[(\beta_1k+\beta_2k^3)-\sqrt{(\beta_1k+\beta_2k^3)^2+\gamma^2m^2}]/\gamma m$. Both $c_{\mathbf{k}\uparrow}$ and $c_{\mathbf{k}\downarrow}$ could hop to $\epsilon_-$ band, with the hopping amplitude $t_{\mathbf{k},\uparrow}=a_kt$ and $t_{\mathbf{k},\downarrow}=b_kt$, respectively. The estimated value of $t$ is much larger than $\Delta$ in the SC, in which case the proximity induced gap can be estimated by~\cite{Chung2011}
\begin{equation}
\Delta_{k}^{\text{1D}} = \left\langle\eta_{\mathbf{k}}E_{\mathbf{k}}/|t_{\mathbf{k},\uparrow}|^2\right\rangle_{k_x}.
\end{equation}
Here
$\eta_{\mathbf{k}}=\langle t_{\mathbf{k},\uparrow}t_{-\mathbf{k},\downarrow}-t_{\mathbf{k},\downarrow}t_{-\mathbf{k},\uparrow}\rangle\Delta/2E_{\bf k}$,  $E_{\bf k}=\sqrt{\Delta^2+(\epsilon_{\bf k}'-\mu')^2}$, $\langle...\rangle_{k_x}$ is averaging over $k_x$, and $k_y$ is taken to be the momentum at the Fermi surface. With NbS$_2$ SC gap 6.5~K, we estimate $\Delta_{k}^{\text{1D}}\approx1.47$~K which is observable experimentally.

{\it Experimental proposal.--}To verify our proposal, we would like to propose an experimental consequence that is simple to measure. We consider a MoS$_2$ flake on top of the superconducting substrate, as is shown in Fig. 4. Due to the crystal symmetry, a corner on such a flake can naturally have a 120$^\circ$ angle. From the crystal structure (see also Fig. 1 (d)) one can see that a Mo-terminated zigzag edge is always adjacent to a S-terminated zigzag edge at a 120$^\circ$ corner. Since we have discussed that the S-terminated edge is almost fully spin-polarized and cannot be coupled to the $s$-wave superconductor, we conclude that every corner of the MX$_2$ monolayer with an angle of 120$^\circ$ is a boundary between TSC and trivial $s$-wave superconductors, if the Mo-terminated edge is brought to the topological superconducting phase. Therefore one expects to see a Majorana zero mode at each such corner, which can be easily verified by scanning tunneling microscope (STM) or transport measurements. The Majorana nature of the zero mode can be further verified by comparing 120$^\circ$ corner with 60$^\circ$ corner. The latter is a border between two topologically equivalent superconductors, which does not have a Majorana zero mode.

This work is supported by the Defense Advanced Research
Projects Agency Microsystems Technology Office, MesoDynamic Architecture Program (MESO) through contract numbers
N66001-11-1-4105 (G. X., J. W. and X. L. Q.) and the European Research Council Advanced Grant (ERC 291472) (B. Y.). G.X. would also like to thank for the support from NSF of China. After finishing this work, we become aware of a similar proposal~\cite{chu2013}. However, the edge state band structure obtained there is different from our {\it ab initio} calculation results.

\end{document}